&latex209
\documentstyle[psfig,subfigure]{mn}

\begin{document}
\newcommand{\goodgap}{%
\hspace{\subfigtopskip}%
\hspace{\subfigbottomskip}}

\title[Iron Needles in Supernova Remnants?]{Iron Needles in Supernova Remnants?}
\author[H.Gomez et al.]
{Haley L. Gomez (n\'{e}e Morgan) $^1$, Loretta Dunne $^2$, Stephen A.
Eales $^1$, \cr Edward L. Gomez $^1$,  Michael G. Edmunds $^1$\\
$^1$Department of Physics
\& Astronomy, Cardiff University, 5 The Parade, Cardiff CF24 3YB,
UK.\\
$^2$School of Physics \& Astronomy, University of Nottingham,
University Park, Nottingham, NG7 2RD, UK.
}
\maketitle

\begin{abstract}
It has been suggested by Dwek (2004) that iron needles could explain
the submillimetre emission from the Cas A supernova remnant (SNR) with only
a very small total mass.  We investigate whether a similar model holds
for the Kepler supernova remnant, and find that its emission could
indeed be explained by a dust mass of less than $\rm
10^{-2}M_{\odot}$, dependent on the axial ratio $l/a$ of the needles -
which we constrain to be less than 700.  But the implied needle model
for Kepler is inconsistent with that suggested for Cas A since {\it
either} the needles would have to have a resistivity one or two orders
of magnitude greater than those in Cas A {\it or} the electron density
in Kepler's shocked plasma must be 40 times greater than suggested by
X-ray observations.  An additional problem with the needle model is
that the implied thickness of the needles seems to be implausibly
small, if the emission properties are calculated under the usual
approximations.

\end{abstract}

\begin{keywords}
ISM -- supernovae:
individual: Kepler, extinction -- submillimetre.
\end{keywords}

\section{Introduction}

Recent observations of young supernova remnants (SNRs) with the
Submillimetre Common User Bolometer Array (SCUBA) at the James Clerk
Maxwell Telescope in Hawaii provided the first evidence for large
quantities of cold dust manufactured in the explosion (Dunne et al.
2003 - hereafter D03), Morgan et al. 2003 - hereafter M03).  This
result provides a plausible explanation for the origin of dust in high
redshift quasars and submillimetre (submm) galaxies
(e.g. Bertoldi et al. 2003, Eales et al (2003), Smail, Ivison
\& Blain 1997). However, there are two conflicting theories which suggest
that the dust mass in the remnants is in fact much lower: (i) the
emission in Cassiopeia A is contaminated with foreground material and
is not associated with the remnant and (ii) needle-like metallic
grains are responsible for the submm emission.  CO observations
suggest that some of the dust in Cas A is from an intervening
molecular cloud.  This cannot be the case for
Kepler which is relatively free of foreground material. However, the
emission from iron needles fits the infrared-submm Spectral Energy
Distribution (SED) of the supernova.  The needles were originally
conceived to provide a mechanism by which to thermalise the microwave
background without a Big Bang model (Hoyle \& Wickramasinghe 1988;
Wright 1982).  Iron needles are formed via a `screw dislocation'
whereby an initially spherical condensation grows exponentially along
one axis.  Such grains can form with axial ratios ($l/a$) as high as
$10^5$.  They produce little extinction in the optical, yet their far
IR-submm opacity is as much as $10^6$ times higher than `normal'
interstellar silicates for $l/a > 1000$ (Hoyle \& Wickramsinghe 1988,
Wright 1982).  In the energetic environment of a SNR, the needles
would be heated by the X-ray plasma to about 10-20 K, consistent with
the SED determined for young SNRs (D03; M03). In this scenario, iron
needles produce the submm peak in the SED while the mid IR emission is
due to collisionally heated `normal' dust.  The mass of iron needles
required to produce the submm flux in Cas A is only $10^{-5}~ \rm
M_{\odot}$ (Dwek 2004), questioning of the origin of dust once more.

In this paper, we address the possibility that iron needles could be
responsible for the emission in Kepler's SNR following the argument in
Dwek (2004) and determine the properties of such dust grains.  In
Section 2 we conclude that if iron needles exist in Kepler's SNR, they
must have different properties to those argued to be in Cas A. Indeed,
we find that the implied physical thickness of the needles seems
implausibly small, although this cannot yet be claimed as a decisive
argument since it may reflect difficulties with the usual
approximations made in calculating the radiation properties of thin
cylinders.

\section{Emission from Iron Needles}
\label{sec:iron}

\begin{figure}
\begin{center}
\psfig{file=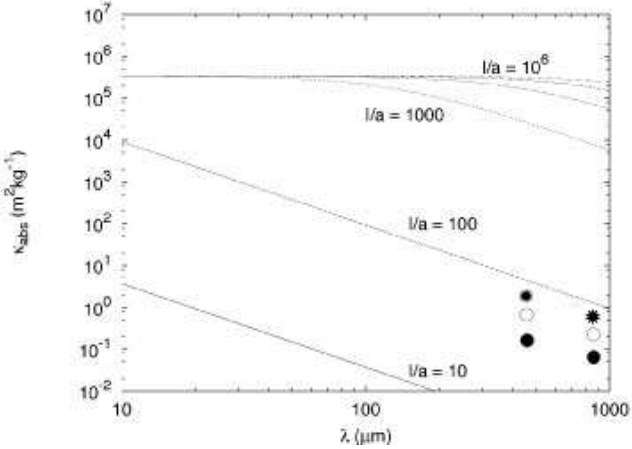,width=8.5cm,height=6cm}
\end{center}
\caption{\small{The mass absorption coefficient of iron
needles for $\rho_r = 4\times 10^{-18} ~\rm{s}$. The axial ratios are
indicated next to each line from 10 to 10,000. For comparison, the
values of $\kappa$ at 450 and 850 $\mu$m for `normal' dust grains are
indicated by the symbols, where asterisks represent $\kappa_{450} =
1.5\rm kg^{-1}m^2$, $\kappa_{850} = 0.76\rm kg^{-1}m^2$ (laboratory
studies); open circles - $\kappa_{450} = 0.88\rm kg^{-1}m^2$,
$\kappa_{850} = 0.3\rm kg^{-1}m^2$ (evolved stars, reflection nebulae
and cold clouds); and filled circles - $\kappa_{450} = 0.26\rm
kg^{-1}m^2$, $\kappa_{850} = 0.07\rm kg^{-1}m^2$ (diffuse ISM).}}
\label{antenna}
\end{figure}

The iron needles proposed by Dwek (2004) are highly elongated.  No
detailed models have yet been made to demonstrate the feasibility of
growing grains with axial ratios of up to 10,000 in the conditions
following a supernova explosion. If such grains do condense then their
metallic properties combined with their extreme elongation make them
highly efficient absorbers and emitters of submm radiation. The dust
mass absorption coefficient for iron needles is given by
\begin{equation}
\kappa_{\rm{abs}}= \frac{4\,\pi}{3c\rho_d \, \rho_r}
\end{equation}
where $\rho_d$ is the density of the iron grains $\sim 8000$ kg
m$^{-3}$ and $\rho_r$ is the electrical resistivity, the value of
which is temperature dependant as well as being very sensitive to the
level of impurity in the grains. Typical values for $\rho_r$ at the
low temperatures found in the ISM are in the range $10^{-17} -
10^{-18}$ s for a reasonable level of impurities (0.5 -- 1 per cent)
(Hoyle \& Wickramasinghe 1988; Dwek 2004 uses units
of $\Omega$ cm for the resistivity where $1
\,\Omega\,\rm{cm} = 1.139\times
10^{-12}~\rm{s}$). $\kappa_{\rm{abs}}$ for iron needles is
approximately $\sim \rm 10^5 - 10^6~m^2 kg^{-1}$. These values are
much higher than those in the literature for dust observed in
astrophysical environments (e.g. ISM, circumstellar, reflection
nebulae) as also shown in Fig. 1.  It is clear that iron needles are
predicted to be incredibly efficient radiators in the submm.  There is
a long wavelength cut-off in the behaviour of the absorption
coefficient (Wright 1982)
\begin{equation}
\lambda_0 = {1
\over{2}}\rho_r c {(l/a)^2\over{\rm{ln}(l/a)}}
\end{equation}
where at $\lambda >
\lambda_0$, $\kappa
\propto (\lambda/ \lambda_0)^{-2}$. The absorption coefficient for
needles of various axial ratios is shown in Fig.~\ref{antenna} for
$\rho_r = 4\times 10^{-18}$ s (following Dwek 2004), where the most
elongated needles have absorption coefficients which are independent
of wavelength throughout the FIR/submm region of the
spectrum. Depending on the combination of parameters, $\kappa$ can be
as much as $10^5$ times greater than the highest value for `normal'
grains estimated from the literature.

As shown by Dwek (2004) the FIR/submm SED of Cas A can be fitted
equally well by a combination of `normal' dust at hot temperatures and
iron needles at $\sim 8$ K. In this case, a much smaller mass of dust
is required due to the high mass absorption coefficient of the
needles. The high emission efficiencies of the iron needles
allow them to reach such low equilibrium temperatures despite the
presence of the hot X-ray plasma. Here, we apply the same analysis to
Kepler's SNR. We take the same range of electrical resistivity as
fitted by Dwek (2004) for Cas A ($2.3\times 10^{-18} -
6\times 10^{-18}$ s) and fit the SED of Kepler with iron needles with
a range of axial ratios. We find acceptable fits for needles with
axial ratios up to $\sim 5000$ and temperatures between 28 -- 10
K. The mass of iron needed depends on the axial ratio (as this affects
the value of $\lambda_0$ and hence $\kappa$ at submm wavelengths) and
ranges from $10^{-7}\,\rm{M_{\odot}}$ for the longest needles with
$l/a \sim 5000$ to $0.8 \,\rm{M_{\odot}}$ for needles with $l/a =
100$. In order to have the minimum temperature of 10 K, we find that
the needles must have $l/a \leq 1000$. The needles are heated by
collisions with electrons and ions in the plasma, and following Dwek,
the heating rate from collisions is given by
\begin{equation}
H_d=2\pi a l n_e(1-f_r)\left[\int^{\infty}_{0} g(E)v(E) E_{dep}(E)dE \right]
\end{equation}
where $n_e$ is the electron density, $f_r$ is the fraction of
electrons that are reflected from the surface of the grain and $g(E)=
2\pi^{-1/2}(kT)^{-3/2}E^{1/2}exp(-E/kT)$ is the Maxwell-Boltzmann
distribution of electron energies with velocity $v(E)$. $E_{dep}$ is
the amount of energy deposited in the neddle by the incident electrons.  

\begin{figure}
\begin{center}
\psfig{file=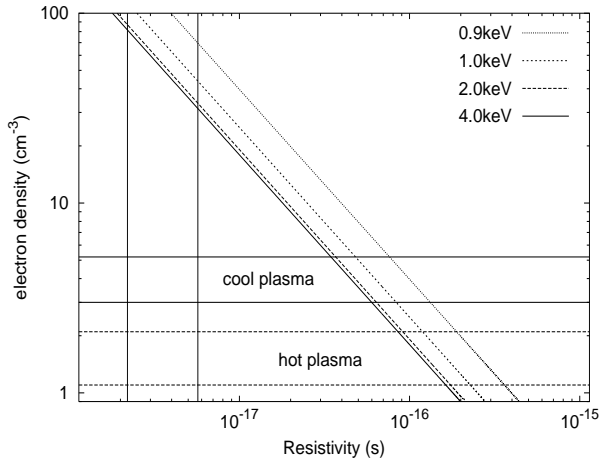,width=8.2cm,height=6.2cm}
\end{center}
\caption{\small{Electron density
required to heat a needle of 10 K in Kepler for different plasma
temperatures with $\lambda_0 = 400~\mu$m.  The diagonal lines show the
variation of $n_e$ with resistivity for constant gas electron
temperature in decreasing order from left to right.  The hot and cool
plasmas are defined by the regions inside the horizontal lines $\rm
\{n_e, T_e\} = \{4.0 \pm 1.0 ~cm^{-3},~ 0.80 \pm 0.2~ keV\}$ and $\rm
\{1.6 \pm 0.5 ~cm^{-3},~ 3.0 \pm 0.9~ keV\}$ from Kinugasa \& Tsunemi
(1999) with $30$ per cent errors. The vertical lines represent the
range of resistivities in Cas A (Dwek 2004) with needles of
8.2 K.}}
\label{dwekfig4_kep}
\end{figure}

The cooling rate is given by the luminosity of the grains  
\begin{equation}
L_d = 4m_d\int\kappa(\lambda)\pi B(\lambda, T_d)d\lambda
\end{equation}
The electron density required for the different plasma temperatures to
heat a needle to 10 K is obtained by equating the heating and cooling
rates ($H_d = L_d$).  The range of values of $n_e$ expected
in Kepler with a cutoff wavelength $\lambda_0=400~\mu$m is shown in
Figure~\ref{dwekfig4_kep}.

In this case, the needle conductivity needs to be $\rm
\sim (4 - 60) \times 10^{-17}~s $ if they reside in the hot or
cold components of the X-ray gas (which is equivalent to $\rm \sim (4
- 60) \times 10^{-5}~\Omega~\rm cm$).  This requires that iron needles
in Kepler must have a resistivity of one to two orders of magnitude
greater than those in Cas A {\it or} that the electron density in
Kepler must be greater than 40$~\rm cm^{-3}$ (i.e. $40 ~\times$
greater than observations suggest - for further details see Morgan
2004).  This result is the same even with the contamination from
foreground material, since Dwek (2004) used an SED with similar far-infrared
fluxes as the {\it Spitzer} observations.

Using the range of resitivities for Kepler from
Figure~\ref{dwekfig4_kep}, we re-fit the SED to be sure that there is
consistency and the allowed temperatures still range from 28 -- 10 K
as before, though now only needles with $l/a < 700$ provide acceptable
fits. Masses of iron range from $10^{-5} - 0.019\, \rm{M_{\odot}}$ for
$l/a = 700 - 100$. Shorter needles would also fit the SED but would
require more mass in iron. It is very important to note that the mass
of iron required is not at all well determined, and this is also the
case for Cas A - the {\em minimum\/} mass can be specified (the
longest needles which will fit the SED) but once the cutoff wavelength
moves shortward of the submm region there is no distinction in the fit
for smaller axial ratios. For Cas A the minimum iron mass
corresponding to 8 K needles with $l/a = 2000$ is $\sim
10^{-4}\,\rm{M_{\odot}}$ but for needles with $l/a = 100$ this rises
to $15 \rm{M_{\odot}}$ of dust! Given there is a limit to the amount
of iron produced in the SNe (Woosley \& Weaver 1995), a limit can be
put on the axial ratios for Cas A of $700 < l/a < 2000$.

So we have the apparent discreprancy that although needles with
$\rho_r \sim (2 - 6) ~\times 10^{-18}~s$ and $l/a
\sim 100 - 4000$ would account for the SED of Cas A (Dwek 2004), we
find here that the SED of Kepler requires needles with $\rho_r \sim (4
- 60) ~\times 10^{-17}~s$ and $l/a \sim < 700$.

\section{A Difficulty and Discussion}

The approximation used to determine the opacity of iron needles in
Dwek (2004) and the literature (including this work), is only valid
for needles with maximum length given by the Rayleigh criterion,
$$l_{max} = {1\over{20\pi}} (\lambda \rho_r c)^{1/2}$$ (Li 2003).
This condition needs to be satisfied since the absorption of the
needles only becomes efficient enough to provide the opacities when
all elements within the particle radiate in phase with each other
(Wright 1982; Li 2003; Krugel 2003).  Using the range of needle
resistivities and axial ratios estimated for Kepler, the maximum
needle length is $0.05 - 0.2~\mu$m. Since we know the range of $l/a$
required to fit the SED, we can estimate that the Rayleigh criterion
is only satisfied in Kepler's supernova remnant for grain radii of
$\rm 0.8 - 5.7~\AA$.  This is equivalent to approximately a few layers
of iron atoms.  As highlighted in Li (2003), it is extremely difficult
to explain how such small grains would form and exist under
astrophysical conditions.  It may be that the emission/absorption
calculations which do not have to use the above approximations might
allow more physical grains, but we are not at present able to confirm
or refute this.

It is also important to highlight the difficulties involved when
determining the dust mass of more `normal' dust grains in Kepler.
Uncertainties in this case include observational errors in the submm
($\sim$ 30 per cent), inadequate knowledge of dust properties (the
absorption efficiency of the grains, $\kappa$, is the largest source
of error here since we do not know which values are relevant
for supernova dust - D03, M03) and uncertainties in the progenitor
mass of the remnant.  The ambiguity in fitting the infrared-submm SED
to constrain the best fit parameters also introduces large errors
(Gomez et al, submitted).  Taking all of these uncertainties into
consideration, the mass of dust in Kepler is not well determined,
although the submm emission provides strong evidence that there is
significantly more dust associated with the remnant than shown by
infrared observations.

If iron needles are responsible for the emission in supernova remnants
then SNe are not important contributors to the interstellar mass
budget.  However, the existence of iron needles would suggest that the
infrared bright dusty galaxies at high redshifts could be attributed
to smaller amounts of dust; there would no longer be a problem with
massive dusty galaxies in the early Universe (Edmunds \&
Wickramasinghe 1975).  Future polarimtery observations of Cas A
and accurate flux measurements in the wavelength range 150 - 350
$\mu$m should provide observational evidence on whether iron needles
do exist in supernovae, although at present we are inclined to favour
more conventional forms of dust.  

\section*{Acknowledgements}
HLM would like to acknowledge the support of a Research Fellowship
from the Royal Exhibition of 1851. We thank Eli Dwek and Aigen Li for
informative discussions and the referee for insightful comments.

\end{document}